 \documentclass[prl,aps,twocolumn,showpacs]{revtex4}

\usepackage{epsfig}
\usepackage{amsfonts}
\usepackage{amsmath}

\begin{document}

\title{Topologically nontrivial $1/3$-magnetization plateau state in a spin-1/2 trimer chain}

\author{Y. Y. Han$^{1}$}
\thanks{These authors contributed equally to this work.}
\author{B. C. Yu$^{3}$}
\thanks{These authors contributed equally to this work.}
\author{Z. Du$^{1}$}
\author{L. S. Ling$^{1}$}
\author{L. Zhang$^{1}$}
\email{leizhang@hmfl.ac.cn}
\author{W. Tong$^{1}$}
\author{C. Y. Xi$^{1}$}
\author{J. L. Zhang$^{1}$}
\author{T. Shang$^{3,4}$}
\email{tshang@phy.ecnu.edu.cn}
\author{Li Pi$^{1,2}$}
\author{Long Ma$^{1}$}
\email{malong@hmfl.ac.cn}

\affiliation{$^{1}$Anhui Key Laboratory of Low-Energy Quantum Materials and Devices, High Magnetic Field Laboratory, Chinese Academy of Sciences, Hefei 230031, China\\
$^{2}$ Hefei National Research Center for Physical Sciences at the Microscale, University of Science and Technology of China, Hefei 230026, China\\
$^{3}$ Key Laboratory of Polar Materials and Devices (MOE), School of Physics and Electronic Science, East China Normal University, Shanghai 200241, China\\
$^{4}$ Chongqing Key Laboratory of Precision Optics, Chongqing Institute of East China Normal University, Chongqing 401120, China
}

\date{\today}

\begin{abstract}
  Topologically nontrivial Haldane phase is theoretically proposed to be realized in the 1/3-magnetization ($M$) plateau of spin-1/2 trimer systems. However, the spin excitation gap, typical characteristic of Haldane phase, is not yet experimentally verified. Here, we report the nuclear magnetic resonance investigations into the low-energy spin dynamics in the $S=1/2$ spin-trimer antiferromagnetic chain compound Na$_2$Cu$_3$Ge$_{4-x}$Si$_{x}$O$_{12}$ ($x=0, 0.1\sim1.5$). In the parent compound ($x=0$), the spin-lattice relaxation rate (1/$T_1$) shows significantly different temperature dependence when the external magnetic field is increased above the critical field of  $\mu_0$$H_{c}$ = 29 T.
  The spin excitation gap is evidenced from the thermally activated behavior of $1/T_1(T)$ in the 1/3-$M$ plateau state. By substituting Ge$^{4+}$ with Si$^{4+}$, the critical field for the 1/3-$M$ plateau significantly decreases, e.g. $\mu_0H_{c}=17$ T in $x=1.0$ samples, which results from the suppressed inter-trimer coupling $J_2$. The gapped spin excitation is confirmed again above 17 T, whose size shows temperature-dependent behavior for $\mu_0H\geq25.72$ T. These observations provide further insights into the Haldane physics.
\end{abstract}

\maketitle

Topology has become an important concept in physics since the discovery of quantum spin Hall effect and topological insulators\cite{Hasan_RMP_2010}.
It supplies a new perspective on states of matter beyond the local order parameters in conventional Ginzburg-Landau theory.
From the application standpoint, as a robust physical property against disturbance of circumstance, topological nontrivial states serve as a robust physical resistant to the circumstance disturbance, supplying the hardware basis for quantum computation with low error rates\cite{Bartlett_PRL_105_110502}. In the fruitful field of topological insulator, the research starts from the band theory based on single particle approximation. While, in strongly-correlated electron systems, exploring topological nontrivial states is still challenging\cite{Wang_Science_343_629,Chen_PRB_83_035107,Pollmann_PRB_85_075125}.

Due to the enhanced spin fluctuations as well as reduced dimensionality, quantum spin chains provide a valuable playground for studying exotic quantum phases, transitions, and emergent spin excitations. As indicated by the Bethe ansatz solution\cite{Bethe_ZPhys_71_205}, the ground state of one-dimensional(1D) spin-1/2 Heisenberg spin chain is the quantum critical Tomonaga-Luttinger liquid. The spin excitation spectrum is dominated by the gapless spinon continuum with a fractional quantum number. While for the integer spin case, a finite spin excitation gap exists, as conjectured by Haldane\cite{Haldane_PLA_93_464,Haldane_PRL_50_1153}. Subsequently, Affleck, Kennedy, Lieb, and Tasaki (AKLT) propose a model Hamiltonian for spin-1  by artificially introducing the quadratic term of spin exchange interactions\cite{Affleck_PRL_59_799}, which can be solved exactly. A novel topological string order and edge states have been found to exist in this AKLT state\cite{Affleck_PRL_59_799}, which is believed to capture the nature of Haldane phase. Further theoretical work reveals that Haldane chains with odd integer spin host typical symmetry protected topological order\cite{Gu_PRB_80_155131,Pollmann_PRB_85_075125}. The Haldane gap and edge states are verified by extensive experimental study of nickel-based materials with spin-1 magnetic sites.

The Haldane phase also can be realized in the magnetization ($M$) plateau state of spin-1/2 trimer chains. Theoretically, the precondition for the existence of plateau in the $M(H)$ curve for quantum spin chains is the Oshikawa-Yamanaka-Affleck rule\cite{Oshikawa_PRL_78_1984},$S\cdot n(1-M/M_{\rm{sat}})=\rm{integer}$.
Here, $S$ and $n$ denote the spin value of the magnetic ion and number of magnetic ions per unit cell.
The $M/M_{\rm{sat}}$ represents the ratio between the magnetization and the saturated magnetization.
The 1/3-$M$ plateau state of spin-1/2 trimmer antiferromagnetic chains is proposed theoretically to be topological nontrivial with a finite Haldane gap\cite{Hu_PRB_90_035150}.

In the spin trimmer chain compound Cu$_3$(P$_2$O$_6$OH)$_2$ ($J_1-J_2-J_2$), a 1/3-$M$ plateau was observed above the critical field of 12 T\cite{Hase_PRB_73_104419}.
A singlet-triplet-like excitation gap responsible for the plateau was observed from the inelastic neutron scattering under zero magnetic field\cite{Hase_PRB_76_064431}.
Recently, emergent composite excitations of doublons and quartons have been reported in another spin-trimmer antiferromagnet Na$_2$Cu$_3$Ge$_4$O$_{12}$, where the 1/3-$M$ plateau appears above $\mu_0H_{c}\sim28$ T\cite{Bera_NC_13_6888}.
However,  to the best of our knowledge, the experimental evidence supporting the existence of a finite Haldane gap in the magnetization plateau state is still missing in the spin-1/2 trimmer chains, clearly requires further investigations.

In this paper, we use nuclear magnetic resonance (NMR) technique to explore the low-energy spin excitations of the magnetization plateau state in Na$_2$Cu$_3$Ge$_{4-x}$Si$_{x}$O$_{12}$ ($x=0, 0.1\sim1.5$). In the parent compound ($x=0$), the incommensurate magnetic ordering is evidenced by the spectral intensity modulation below $T_N=2$ K in the powder samples aligned by magnetic field. The 1/3-$M$ plateau state appears above
$\mu_0H_{c}=29$ T, confirmed by high-field magnetization measurements. A thermally activated behavior is observed from the temperature dependent nuclear spin-lattice relaxation rates. This clearly indicates the gap opening in the spin excitation spectrum, which is believed to be the topologically nontrivial Haldane gap. We further tune the magnetic coupling between the spin trimmers via replacing Ge$^{4+}$ with smaller Si$^{4+}$.
Upon doping, the magnetic transition is quickly suppressed and the critical field above which the spin gap emerges shifts to the lower field side. This effect is most likely attributed to the reduced inter-trimer magnetic coupling. For $x=1.0$ samples under magnetic fields ranging from 25.72 T to 32.5 T, the spin gap size obtained from the spin-lattice relaxation rates varies for different temperature region, which can be attributed to the dominant spin excitations at different wave vector $\bf{q}$ in the reciprocal space.

Polycrystalline samples of Na$_2$Cu$_3$Ge$_{4-x}$Si$_{x}$O$_{12}$ were synthesized using a solid-state reaction method\cite{Bera_NC_13_6888}.
The NMR measurements were performed on $^{23}$Na nuclei ($\gamma_n=11.262$ MHz/T, $I=3/2$) with a phase-coherent NMR spectrometer.
The spectrum was obtained by summing up the spectral intensity with the frequency swept.
The spin-lattice relaxation rate ($1/T_1$) was measured with the inversion-recovery sequence, and obtained by fitting the time dependence of nuclear magnetization to ($M(t)$) to $M(t)/M(\infty)= 1-0.1\exp[-(t/T_1)^{\beta}]-0.9\exp[-(6t/T_1)^{\beta}]$.

\begin{figure}
\includegraphics[width=8cm, height=6cm]{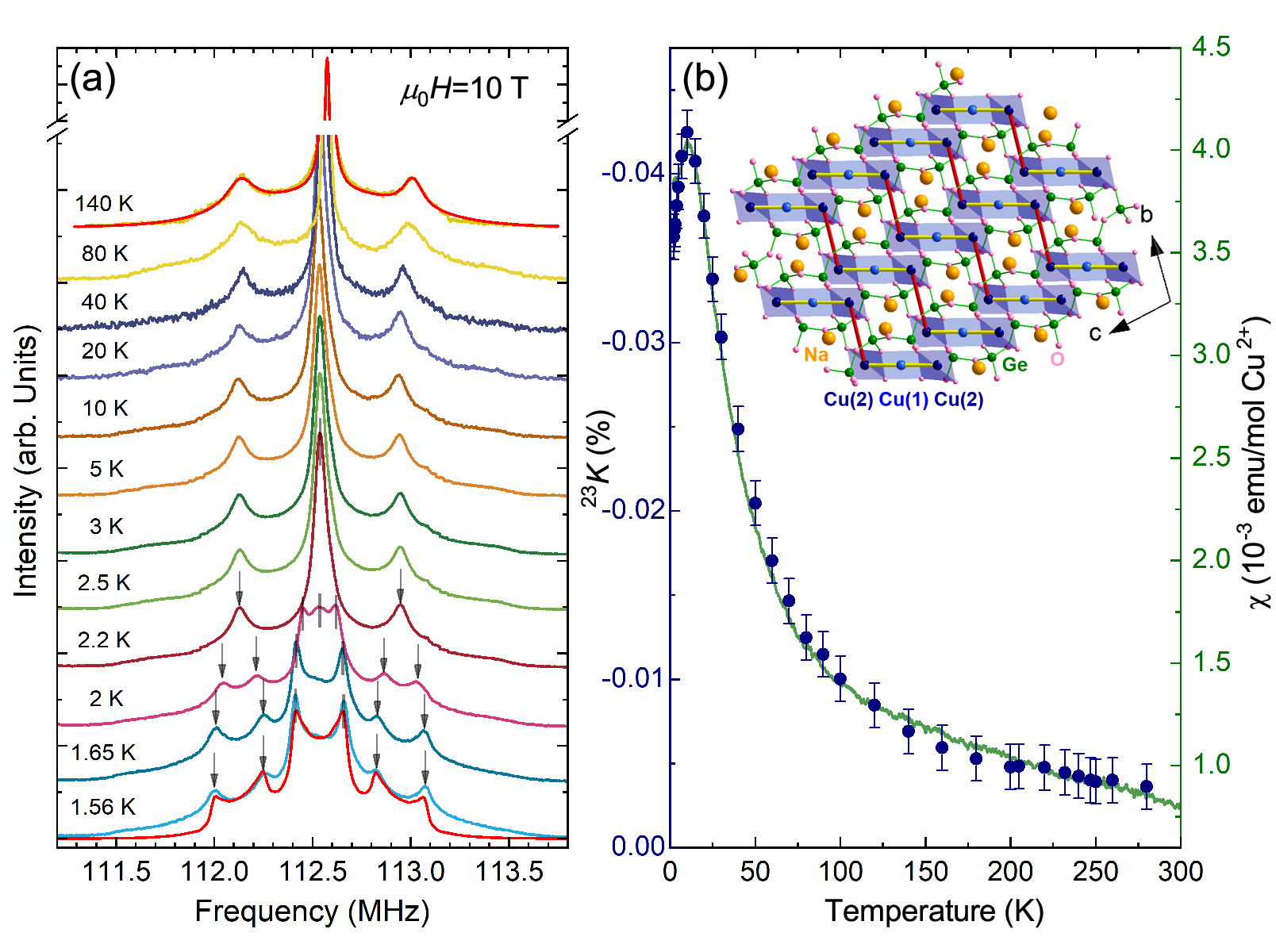}
\caption{\label{spec1}(color online) (a) $^{23}$Na NMR spectra at different temperatures under a 10 T magnetic field. The powder sample was finely grounded, loosely put into a capsule and aligned with the applied NMR magnetic field. The red lines are simulations to the spectra.
(b) Temperature dependence of $^{23}$Na Knight shifts under 10 T and $dc$-susceptibility. The $dc$-susceptibility was measured under a weak magnetic field of 0.1 T. The crystal structure of Na$_2$Cu$_3$Ge$_4$O$_{12}$ is sketched in the inset.
}
\end{figure}

The Na$_2$Cu$_3$Ge$_{4-x}$Si$_{x}$O$_{12}$ samples crystallize in a triclinic structure with the $P\text{-}1$ space group\cite{Bera_NC_13_6888}, as shown in the inset of Fig.\ref{spec1}(b).
Three edge-sharing CuO$_4$ planes form Cu$_3$O$_8$ spin trimers with a intra-trimer coupling $J_1$ (shown by yellow lines).
The Cu$_3$O$_8$-trimers interact with each other via O-Ge-O super-superexchange interactions, whose coupling strength is denoted as $J_2$ (shown by red lines). For each trimer, the Cu$^{2+}$ on both sides couple via oxygen-oxygen ions, resulting in the magnetic coupling of $J_3$.
The magnetic transition at $T_N=2$ K results from the weak inter-chain coupling $J_4$, bridged by the Na$^+$ and Ge$^{4+}$ ions.

Microscopic information about the magnetic order can be obtained from the NMR spectrum of the aligned sample. Above $T_N$, $^{23}$Na spectrum is composed by one central peak and double satellite peaks. We simulate the spectrum at $T=140$ K with three lorentz peaks and one gaussian background counting for the deviation from complete alignment. This is consistent with the single Wyckoff position of Na$^+$ ions carrying nuclear spin $I=3/2$.
For temperatures below $T_N$, each peak broadens and exhibits a distinctive double-horn shape, expected for the incommensurate magnetic order\cite{Zeng_CPL_39_107501,Higa_PRB_96_024405}.
The resonance line shape for the incommensurate modulation is simulated as shown by the red line at the bottom of Fig.\ref{spec1}(a), consistent with the previous NMR study\cite{Yasui_JAP_115_17E125}. The Knight shifts, measuring the spin susceptibility, exhibit a highly similar behavior to the temperature-dependent $dc$-susceptibility (See Fig.\ref{spec1}(b)). The broad maximum at $T^*\sim11$ K indicate the setup of short-range order upon cooling, which is commonly observed in low-dimensional magnets\cite{Bonner_PR_135_A640}.

\begin{figure}
\includegraphics[width=8cm, height=11.2cm]{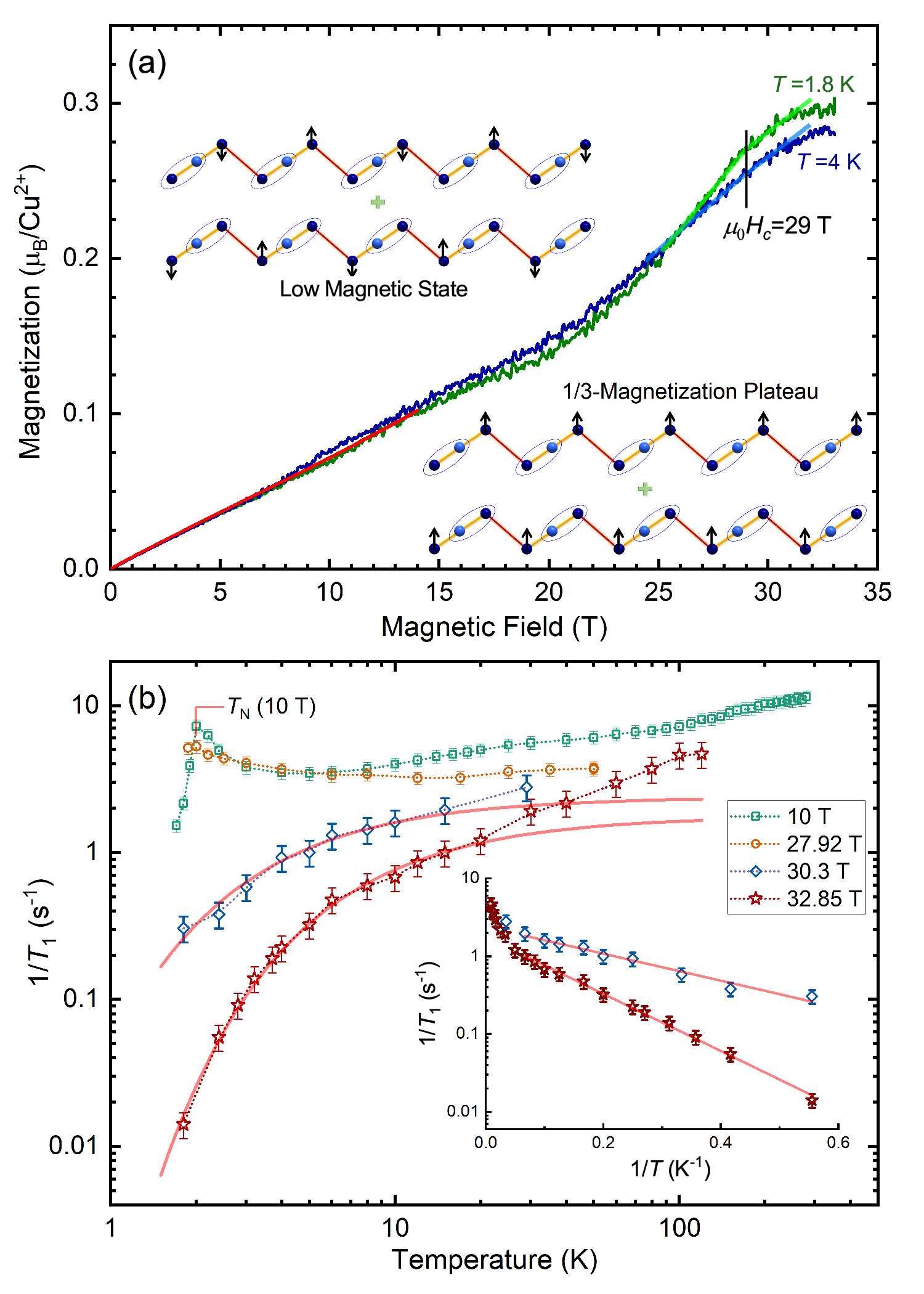}
\caption{\label{SLRR2}(color online)
  (a) Field dependence of the magnetization at $T=1.8$ K and $T=4$ K in Na$_2$Cu$_3$Ge$_4$O$_{12}$.
  The sketched spin configurations of the low magnetic state and plateau state are shown as insets.
  (b) Spin-lattice relaxation rates as a function of temperature under magnetic fields ranged from 10 T to 32.85 T.
  Inset: Semi-log plot of the $1/T_1$ versus the inverse of temperature.
  Both the red lines in the main figure and inset denote the thermally activated behavior of $1/T_1(T)$.
}
\end{figure}

Fig.\ref{SLRR2}(a) presents the $dc$-magnetization versus external magnetic field measured at $T=1.8$ K and 4 K. The magnetization first shows a linear field dependence at $\mu_0H<22$ T, then increases rapidly up to $\mu_0H_{c}=29$ T, followed by a level-off behavior under higher field. The slope change point is defined as the critical field for the 1/3-$M$ plateau state, which is close to the $\mu_0H_{c}=28$ T obtained by previous magnetization measurements under pulsed magnetic field. The corresponding spin states in different field range are depict in the inset of Fig.\ref{SLRR2}(a). The circled dumbbells represent the spin singlets formed by two adjacent Cu$^{2+}$. The moments of the remaining Cu$^{2+}$ ion in the spin trimmer is denoted by black arrows, which are polarized in the plateau.

To verify the existence of spin gap in the magnetization plateau state, we measured the temperature dependence of the spin-lattice relaxation rate under various magnetic fields (see Fig.\ref{SLRR2}(b)). For $\mu_0H=10$ T, the $1/T_1$ first decreases with cooling the sample, then increases at $T_N<T<5$ K, and finally drops steeply below $T_N$. The peak centered at $T_N$ results from the critical slowing down of electron spin dynamics near the transition temperature. At $\mu_0H=27.92$ T just below $\mu_0H_{c}$, the upturn behavior persists down to the lowest temperature, indicating strong spin fluctuations near the magnetic instability. When the field intensity increases to 30.3 T, just 1.3 T higher than $\mu_0H_{c}$, a thermally activated behavior dominates the low temperature $1/T_1(T)$ instead, in sharp contrast with that below $\mu_0H_{c}$. For higher $\mu_0H=32.85$ T, the $1/T_1(T)$ drops more steeply upon cooling, demonstrating a larger spin excitation gap.
In the inset, we present the semi-log plot of $1/T_1$ versus inverse temperature. The gapped behavior is shown more clearly by the straight lines, whose slopes are directly proportional to the gap size.

\begin{figure}
\includegraphics[width=8cm, height=6.1cm]{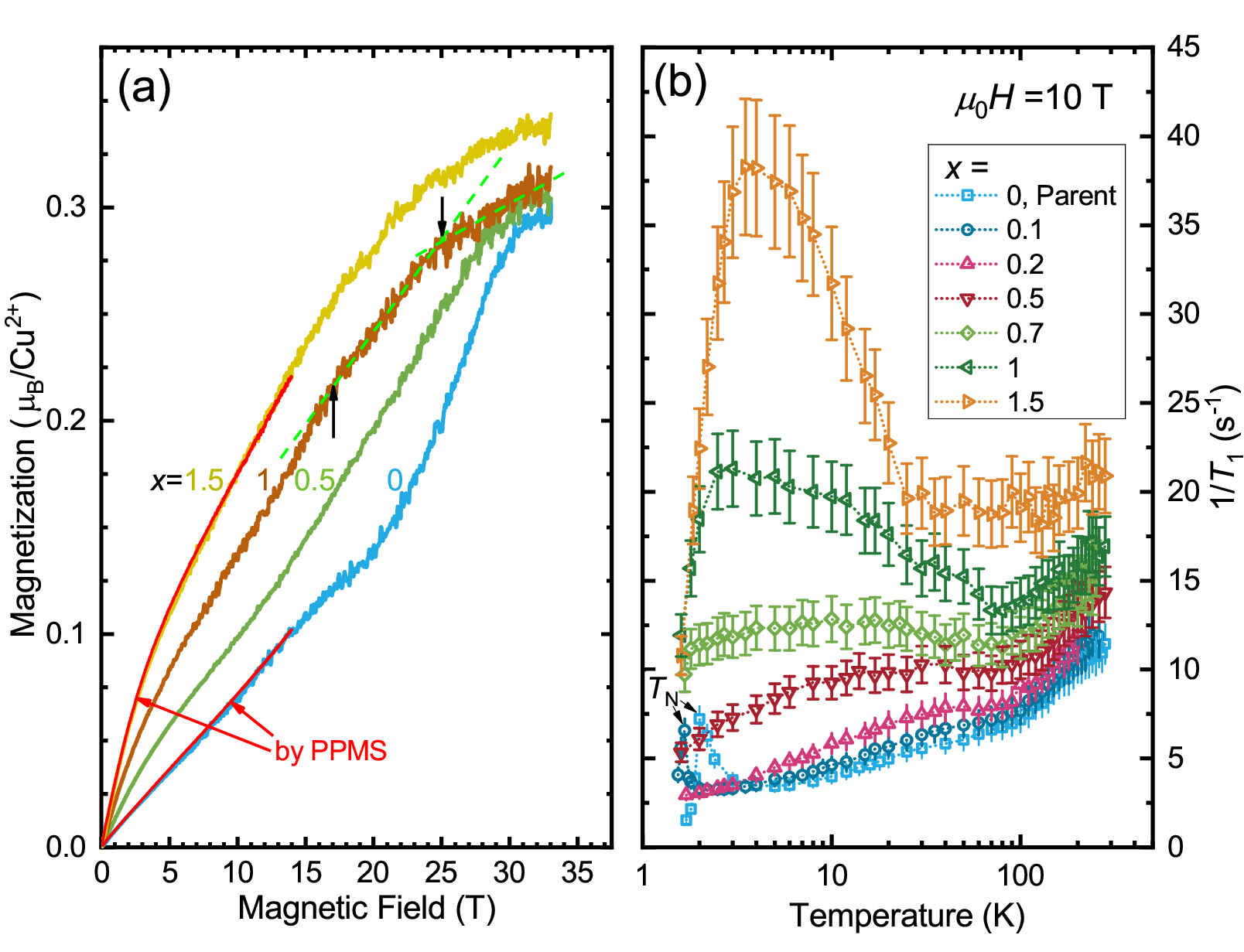}
\caption{\label{doping3}(color online)
  (a) The magnetization at $T=1.8$ K as a function of field intensity in Na$_2$Cu$_3$Ge$_{4-x}$Si$_{x}$O$_{12}$ with $x=$0, 0.5, 1, and 1.5.
  The red lines represent the magnetization measured by the VSM option of PPMS below $\mu_0H=14$ T.
  (b) Temperature dependence of the $^{23}$Na spin-lattice relaxation rate under $\mu_0H=10$ T in samples with $x=$0, 0.1, 0.2, 0.5, 0.7, 1, and 1.5.
}
\end{figure}

The critical field for the magnetization plateau state in spin trimmer systems is closely related to the ratio of inter-trimer to intra-trimer magnetic coupling strength\cite{Bera_NC_13_6888}. We take a step forward to tune the magnetic coupling and thus, the critical field via chemical doping. The interaction between spin trimmers is bridged by the oxygen, germanium and oxygen. By substituting the Ge with Si, we expect effective modification to the inter-trimer coupling $J_2$. As shown in Fig.\ref{doping3}(a), low temperature field-dependent magnetization was measured for Na$_2$Cu$_3$Ge$_{4-x}$Si$_x$O$_{12}$ with various Si contents. The predominant characteristic of the $M(H)$ curve is maintained upon doping except two facts. First, for higher doping levels, the magnetization below 20 T enhances more rapidly with increasing magnetic field.
The level-off trend appears at much lower critical fields upon doping. Second, for $x=1$ sample, the $M(H)$ curve bends over at a moderate field $\mu_0H_{c}=17$ T before its flattening at $\mu_0H=25$ T, which are marked by the up and down arrows. The level-off behavior becomes gradual in the $x=1.5$ sample, which makes the determination of $H_{c}$ difficult. The decreasing $H_{c}$ in doped samples is consistent with the reduced $J_2$ according to the theoretical phase diagram\cite{Bera_NC_13_6888}.

The temperature dependence of spin-lattice relaxation rates in doped samples is measured under a constant magnetic field of 10 T (see Fig.\ref{doping3}(b)). The relaxations of $^{23}$Na nuclei overall get faster with the doping level increasing. The $1/T_1$ shares a similar linear temperature dependence at $T>60$ K. When the sample is further cooled down, a low-temperature broad maximum in $1/T_1(T)$ gradually emerges as the doping level increases. For $x=0\sim0.2$, the sharp peak, corresponding to the critical slowing down of spin fluctuations near the magnetic transition, is suppressed to a lower temperature of $T_N=1.67$ K for $x=0.1$, and finally disappears in the $x=0.2$ sample for the present temperature range.

The linear temperature dependence of $1/T_1$ at $T\geq60$ K is contributed by the uniform spin susceptibility as indicated by the scaling theory study of $S=1/2$ Heisenberg antiferromagnetic spin chain systems\cite{Sachedev_PRB_50_13006}. This relation is verified experimentally in various $S=1/2$ spin chains, e.g. Sr$_2$CuO$_3$\cite{Thurber_PRL_87_247202}. The suppression of the magnetic transition indicates the fragility of the magnetic order against chemical substitution in Na$_2$Cu$_3$Ge$_4$O$_{12}$, which may relate with the reduced inter-chain coupling $J_4$ in the doped samples. As indicated by the low temperature $M(H)$ curve shown in Fig.\ref{doping3}(a), no obvious change of slope is observed near $\mu_0H=10$ T, where the magnetization is far from 1/3 of the saturated one. Thus, the spin state at $\mu_0H=10$ T cannot be regarded as that in the plateau state. Based on this consideration, the low temperature broad maximum in the samples of $x=1.0$ and $1.5$ is most likely due to the emergence of certain type of short-range order with typical timescale close to the inverse Larmor frequency ($\sim112.5$ MHz).

\begin{figure}
\includegraphics[width=8cm, height=6cm]{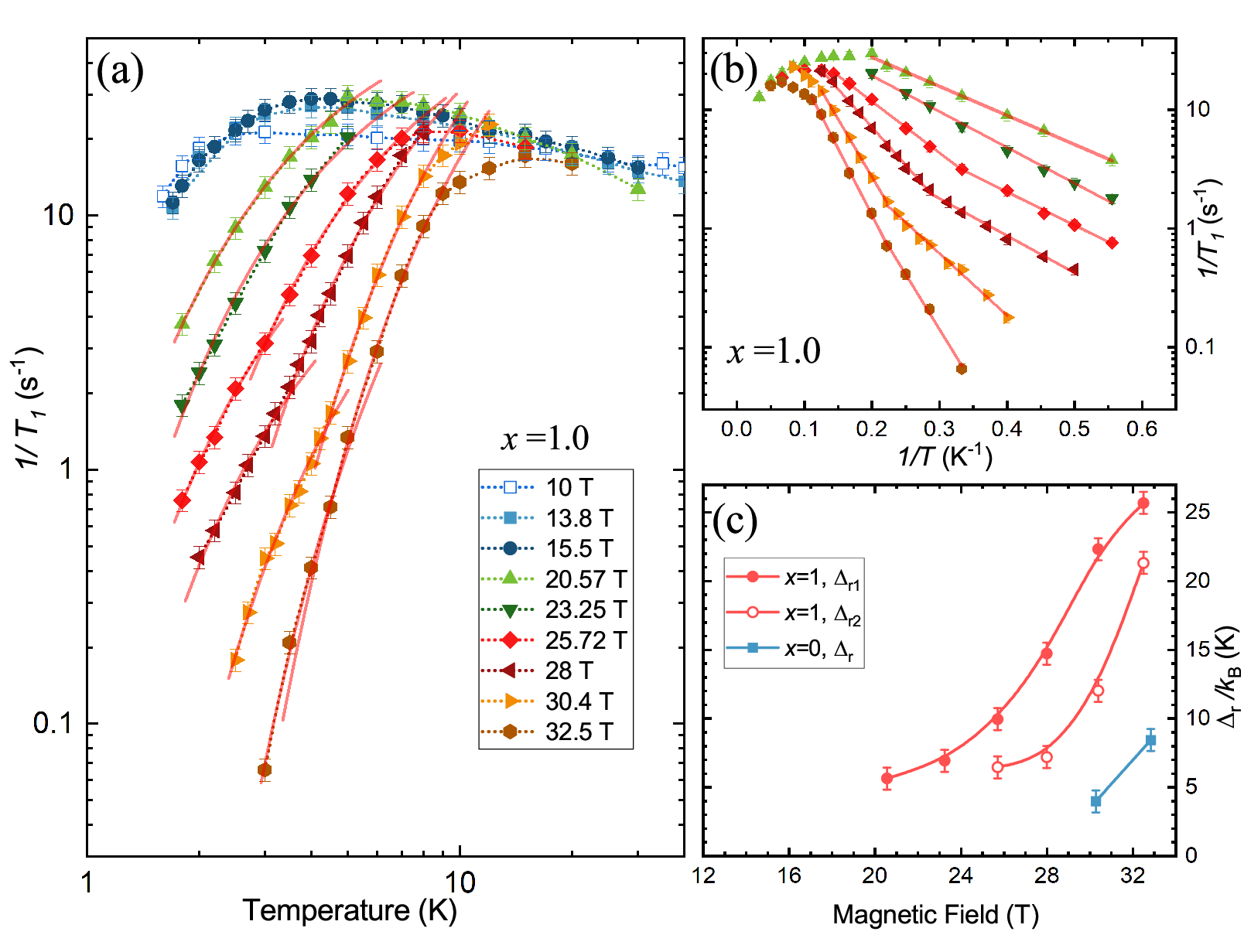}
\caption{\label{gap4}(color online)
  (a) Spin-lattice relaxation rates as a function of temperature in the $x=1.0$ sample with the field intensity ranging from 10 T to 32.5 T.
  (b) The semi-log plot of $1/T_1$ versus the inverse temperature.
  (c) Field dependence of the spin excitation gap size determined by $1/T_1(T)$ in the $x=0$ and $1.0$ samples.
}
\end{figure}

The suppressed critical field for the 1/3-$M$ plateau state in doped samples provides a valuable opportunity to track the field dependence of the spin excitation gap. In Fig.\ref{gap4}(a), we show the temperature dependence of $^{23}$Na spin-lattice relaxation rates under various magnetic fields up to 32.5 T. The low temperature $1/T_1(T)$ drops steeply upon cooling when the field intensity exceeds 17 T. Under higher fields, even more pronounced drop in $1/T_1(T)$ was observed at low temperatures. A kinked characteristic was observed for fields larger than 25.72 T as demonstrated by the thermally activated function shown by the red lines in Fig.\ref{gap4}(a). The gapped behavior can be seen more clearly by the straight lines in the semi-log plot of $1/T_1$ versus the inverse temperature (see Fig.\ref{gap4}(b)). The kinks reflect the change of gap size, which equals to the slopes of straight lines. The field dependence of the spin excitation gap as determined from $1/T_1(T)$ is shown in Fig.\ref{gap4}(c) for the parent sample and sample with $x=1.0$. The $\Delta_{r1}$ and $\Delta_{r2}$ denote the gap size determined from the $1/T_1(T)$ of the $x=1.0$ sample at intermediate and low temperatures, respectively. For the field range in this study, the gap size increases rapidly with magnetic field strength.
The $\Delta_{r1}/\Delta_{r2}$ equals to 1.54, 2.04, 1.86, and 1.2 for $\mu_0H=25.72$ T, 28 T, 30.4 T and 32.5 T, respectively.

Information about the spin dynamics of the magnetization plateau state can be inferred from nuclear relaxation measurements.
The spin-lattice relaxation results from the hyperfine coupling between a nuclear spin and electron spins. The $1/T_1$ can be expressed as a weighted sum over momenta $\overrightarrow{q}$ of the spin-correlation function $S(\overrightarrow{q},\omega_0)$\cite{Moriya_JPSJ_18_516}, i.e.,
$1/T_1=\frac{\gamma_N^2\hbar}{2\mu_B^2}\sum_{\overrightarrow{q}}\lvert A_{\perp}(\overrightarrow{q})\rvert^2S(\overrightarrow{q},\omega_0)$.
Here, $A_{\perp}(\overrightarrow{q})$ denotes the momentum-dependent hyperfine coupling constant in the plane perpendicular to the applied magnetic field. The $\omega_0$ is NMR frequency, generally treated as zero in the studied correlated spin system.

For the Haldane spin chains, the $1/T_1(T)$ at low temperatures is expected to be dominated by a thermally activated behavior with the function form, i.e., $1/T_1\propto\exp(-\gamma\Delta_H/T)$, where $\Delta_H$ is the Haldane gap size, and $\gamma$ is the ratio between the gap $\Delta_r$ determined from $1/T_1(T)$ and $\Delta_H$. The $\gamma$-value varies among different Haldane spin systems. Experimentally, the temperature dependence of $1/T_1$ gives $\gamma=1$ in the typical Haldane spin system Ni(C$_2$H$_8$N$_2$)$_2$NO$_2$ClO$_4$ (NENP)\cite{Gaveau_EPL_12_647}. However, $\gamma$ is determined to be $\sim1.2$ in Y$_2$BaNiO$_5$ and $\sim1.5$ in AgVP$_2$S$_6$, respectively\cite{Shimizu_PRB_52_R9835,Takigawa_PRB_52_R13087,Takigawa_PRL_76_2173}. In the $S=1$ spin chain compound SrNi$_2$V$_2$O$_8$, the $1/T_1(T)$ in the temperature range of 15 K $<T<$ 80 K and 4 K $<T<$ 10 K give very different $\gamma$-values\cite{Pahari_PRB_73_012407}. The behavior sharply contrasts with the temperature dependence of Knight shifts, which measures the spin susceptibility at $\overrightarrow{q}=0$, implying prominent $\overrightarrow{q}$-dependence of the low-energy spin dynamics. Theoretically, the $1/T_1$ can be calculated by assuming a uniform hyperfine coupling constant in the reciprocal space\cite{Jolicoeur_PRB_50_9265,Sagi_PRB_53_9188,Sachdev_PRL_78_943,Konik_PRB_68_104435}. Controversies about the $\gamma$-value still exists depending on the different theoretical approaches.

Recently, the NMR relaxation rate in the Haldane spin chains has been calculated using numerical techniques\cite{Capponi_PRB_100_094411}, by taking into count the contributions from modes with momenta close to $q\approx0$ and $q\approx\pi$. For the low temperature regime ($T<0.41\Delta_H$), the $1/T_1$ is mainly contributed by two-magnon Raman process at $q=0$ as a result of $\omega_0\ll\Delta_H$ and the energy conservation law. The $1/T_1(T)$ follows a simple thermally activated behavior, i.e., $1/T_1\propto\exp(-\Delta_H/T)$. For $0.41\Delta_H<T<\Delta_H$, the spin excitations at $q=\pi$ dominate nuclear relaxations, leading to a steeper drop of $1/T_1$ upon cooling,where the $\gamma$-value is larger than 1.5.

As shown in Fig.\ref{gap4}, the temperature dependence of $1/T_1$ in the $x=1.0$ sample yields distinct spin gap sizes denoted as $\Delta_{r1}$ and $\Delta_{r2}$ for the intermediate and low temperature regime. We understand this phenomenon with the numerical results for $S=1$ Haldane chains as illustrated above. The variation of $\Delta_{r1}/\Delta_{r2}$ is attributed to the fine tuning of the weighted momentum contribution to $1/T_1$ by applied fields.

For both $x=0$ and $x=1.0$ samples, the spin excitation gap size of the 1/3-$M$ plateau state increases with the increasing magnetic field as depicted in Fig.\ref{gap4}(c). For typical $S=1$ Haldane chains, the spin excitation gap decreases with applying magnetic field, and finally closes above the critical field, which results from Zeeman splitting of the three-fold degenerated excited state\cite{Fujiwara_PRB_45_7837,Fujiwara_PRB_47_11860}. In this study, the magnetic field as high as 32.5 T enables us to access the 1/3-$M$ plateau state. However, it is insufficient to fully suppress the gapless spinon excitations observed at zero field\cite{Bera_NC_13_6888}. When the spin-trimer system enters the plateau state from low magnetic field side, the gradually suppressed gapless spinon excitation leads to the increasing gap size observed here. We expect that the spin excitation gap size will decrease under higher magnetic fields.

For isotropic $S=1$ Heisenberg spin chains, the Haldane gap $\Delta_H$ equals to $0.41J$ ($J$ is the exchange interaction constant)\cite{Nightingale_PRB_33_659}. In Na$_2$Cu$_3$Ge$_4$O$_{12}$, the intra-trimer coupling $J_1$ is determined to be $\sim235$ K from numerical simulations of the magnetization data\cite{Bera_NC_13_6888}.
The Cu$_3$O$_8$ spin trimers formed by edge-sharing CuO$_4$ planes remain unchanged upon chemical doping at Ge$^{4+}$-sites. Thus, the Haldane gap for both $x=0$ and $x=1.0$ samples is estimated to be $\sim96$ K. However, the upper limit for the Haldane gap observed from $1/T_1(T)$ under a 32.5 T field is $\sim26$ K, about one quarter of the calculated value. There are two possible reasons for this discrepancy. One is the fore mentioned contribution of the gapless spinon excitations to $1/T_1$, which prevents the determination of intrinsic Haldane gap size. The other one is the
existence of single ion anisotropy, which reduces the Haldane gap size, and is not considered in the proposed spin model.

In summary, we have performed high-field NMR study on the low-energy spin dynamics in the $S=1/2$ spin-trimer chain antiferromagnet Na$_2$Cu$_3$Ge$_{4-x}$Si$_{x}$O$_{12}$ ($x=0\sim1.5$). A non-zero spin gap is verified from the temperature dependence of $1/T_1$ for the 1/3-$M$ plateau state in the parent compound. By suppressing the inter-trimer coupling via silicon doping, the critical field decreases to 17 T for samples with $x=1.0$, which is accessible for commercial superconducting magnets. The gapped spin excitation is confirmed again above 17 T, whose size shows temperature-dependent behavior for $\mu_0H\geq25.72$ T. The temperature dependent gap size can be understood within the numerical calculations of $1/T_1$ contributed by different momenta. The variation of $\Delta_{r1}/\Delta_{r2}$ under different fields is most likely attributed to the fine tuning of the weighted momentum contribution to $1/T_1$ by applied fields. These observations provide direct experimental evidence for the existence of topologically nontrivial Haldane phase in the 1/3-$M$ plateau of $S=1/2$ spin-trimer chain antiferromagnets. The doped samples with lower $H_{c}$ provide a better platform to perform further experimental investigations on the 1/3-$M$ plateau state.

This research was supported by the National Natural Science Foundation of China (Grants No. 11874057, 12374105 and 12374128), the Natural Science Foundation of Shanghai (Grants No. 21ZR1420500 and 21JC1402300), Natural Science Foundation of Chongqing (Grant No. CSTB-2022NSCQ-MSX1678),the Basic Research Program of the Chinese Academy of Sciences Based on Major Scientific Infrastructures (Grant No. JZHKYPT-2021-08) and Fundamental Research Funds for the Central Universities. A portion of this work was supported by the High Magnetic Field Laboratory of Anhui Province.

\begin{thebibliography}{35}
\expandafter\ifx\csname natexlab\endcsname\relax\def\natexlab#1{#1}\fi
\expandafter\ifx\csname bibnamefont\endcsname\relax
  \def\bibnamefont#1{#1}\fi
\expandafter\ifx\csname bibfnamefont\endcsname\relax
  \def\bibfnamefont#1{#1}\fi
\expandafter\ifx\csname citenamefont\endcsname\relax
  \def\citenamefont#1{#1}\fi
\expandafter\ifx\csname url\endcsname\relax
  \def\url#1{\texttt{#1}}\fi
\expandafter\ifx\csname urlprefix\endcsname\relax\def\urlprefix{URL }\fi
\providecommand{\bibinfo}[2]{#2}
\providecommand{\eprint}[2][]{\url{#2}}

\bibitem[{\citenamefont{Hasan and Kane}(2010)}]{Hasan_RMP_2010}
\bibinfo{author}{\bibfnamefont{M.~Z.} \bibnamefont{Hasan}} \bibnamefont{and}
  \bibinfo{author}{\bibfnamefont{C.~L.} \bibnamefont{Kane}},
  \bibinfo{journal}{Rev. Mod. Phys.} \textbf{\bibinfo{volume}{82}},
  \bibinfo{pages}{3045} (\bibinfo{year}{2010}).

\bibitem[{\citenamefont{Bartlett et~al.}(2010)\citenamefont{Bartlett, Brennen,
  Miyake, and Renes}}]{Bartlett_PRL_105_110502}
\bibinfo{author}{\bibfnamefont{S.~D.} \bibnamefont{Bartlett}},
  \bibinfo{author}{\bibfnamefont{G.~K.} \bibnamefont{Brennen}},
  \bibinfo{author}{\bibfnamefont{A.}~\bibnamefont{Miyake}}, \bibnamefont{and}
  \bibinfo{author}{\bibfnamefont{J.~M.} \bibnamefont{Renes}},
  \bibinfo{journal}{Phys. Rev. Lett.} \textbf{\bibinfo{volume}{105}},
  \bibinfo{pages}{110502} (\bibinfo{year}{2010}).

\bibitem[{\citenamefont{Wang et~al.}(2014)\citenamefont{Wang, Potter, and
  Senthil}}]{Wang_Science_343_629}
\bibinfo{author}{\bibfnamefont{C.}~\bibnamefont{Wang}},
  \bibinfo{author}{\bibfnamefont{A.~C.} \bibnamefont{Potter}},
  \bibnamefont{and} \bibinfo{author}{\bibfnamefont{T.}~\bibnamefont{Senthil}},
  \bibinfo{journal}{Science} \textbf{\bibinfo{volume}{343}},
  \bibinfo{pages}{629} (\bibinfo{year}{2014}).

\bibitem[{\citenamefont{Chen et~al.}(2011)\citenamefont{Chen, Gu, and
  Wen}}]{Chen_PRB_83_035107}
\bibinfo{author}{\bibfnamefont{X.}~\bibnamefont{Chen}},
  \bibinfo{author}{\bibfnamefont{Z.-C.} \bibnamefont{Gu}}, \bibnamefont{and}
  \bibinfo{author}{\bibfnamefont{X.-G.} \bibnamefont{Wen}},
  \bibinfo{journal}{Phys. Rev. B} \textbf{\bibinfo{volume}{83}},
  \bibinfo{pages}{035107} (\bibinfo{year}{2011}).

\bibitem[{\citenamefont{Pollmann et~al.}(2012)\citenamefont{Pollmann, Berg,
  Turner, and Oshikawa}}]{Pollmann_PRB_85_075125}
\bibinfo{author}{\bibfnamefont{F.}~\bibnamefont{Pollmann}},
  \bibinfo{author}{\bibfnamefont{E.}~\bibnamefont{Berg}},
  \bibinfo{author}{\bibfnamefont{A.~M.} \bibnamefont{Turner}},
  \bibnamefont{and} \bibinfo{author}{\bibfnamefont{M.}~\bibnamefont{Oshikawa}},
  \bibinfo{journal}{Phys. Rev. B} \textbf{\bibinfo{volume}{85}},
  \bibinfo{pages}{075125} (\bibinfo{year}{2012}).

\bibitem[{\citenamefont{Bethe}(1931)}]{Bethe_ZPhys_71_205}
\bibinfo{author}{\bibfnamefont{H.}~\bibnamefont{Bethe}}, \bibinfo{journal}{Z.
  Phys.} \textbf{\bibinfo{volume}{71}}, \bibinfo{pages}{205}
  (\bibinfo{year}{1931}).

\bibitem[{\citenamefont{Haldane}(1983{\natexlab{a}})}]{Haldane_PLA_93_464}
\bibinfo{author}{\bibfnamefont{F.}~\bibnamefont{Haldane}},
  \bibinfo{journal}{Phys. Lett. A} \textbf{\bibinfo{volume}{93}},
  \bibinfo{pages}{464} (\bibinfo{year}{1983}{\natexlab{a}}).

\bibitem[{\citenamefont{Haldane}(1983{\natexlab{b}})}]{Haldane_PRL_50_1153}
\bibinfo{author}{\bibfnamefont{F.~D.~M.} \bibnamefont{Haldane}},
  \bibinfo{journal}{Phys. Rev. Lett.} \textbf{\bibinfo{volume}{50}},
  \bibinfo{pages}{1153} (\bibinfo{year}{1983}{\natexlab{b}}).

\bibitem[{\citenamefont{Affleck et~al.}(1987)\citenamefont{Affleck, Kennedy,
  Lieb, and Tasaki}}]{Affleck_PRL_59_799}
\bibinfo{author}{\bibfnamefont{I.}~\bibnamefont{Affleck}},
  \bibinfo{author}{\bibfnamefont{T.}~\bibnamefont{Kennedy}},
  \bibinfo{author}{\bibfnamefont{E.~H.} \bibnamefont{Lieb}}, \bibnamefont{and}
  \bibinfo{author}{\bibfnamefont{H.}~\bibnamefont{Tasaki}},
  \bibinfo{journal}{Phys. Rev. Lett.} \textbf{\bibinfo{volume}{59}},
  \bibinfo{pages}{799} (\bibinfo{year}{1987}).

\bibitem[{\citenamefont{Gu and Wen}(2009)}]{Gu_PRB_80_155131}
\bibinfo{author}{\bibfnamefont{Z.~C.} \bibnamefont{Gu}} \bibnamefont{and}
  \bibinfo{author}{\bibfnamefont{X.~G.} \bibnamefont{Wen}},
  \bibinfo{journal}{Phys. Rev. B} \textbf{\bibinfo{volume}{80}},
  \bibinfo{pages}{155131} (\bibinfo{year}{2009}).

\bibitem[{\citenamefont{Oshikawa et~al.}(1997)\citenamefont{Oshikawa, Yamanaka,
  and Affleck}}]{Oshikawa_PRL_78_1984}
\bibinfo{author}{\bibfnamefont{M.}~\bibnamefont{Oshikawa}},
  \bibinfo{author}{\bibfnamefont{M.}~\bibnamefont{Yamanaka}}, \bibnamefont{and}
  \bibinfo{author}{\bibfnamefont{I.}~\bibnamefont{Affleck}},
  \bibinfo{journal}{Phys. Rev. Lett.} \textbf{\bibinfo{volume}{78}},
  \bibinfo{pages}{1984} (\bibinfo{year}{1997}).

\bibitem[{\citenamefont{Hu et~al.}(2014)\citenamefont{Hu, Cheng, Xu, Luo, and
  Chen}}]{Hu_PRB_90_035150}
\bibinfo{author}{\bibfnamefont{H.~P.} \bibnamefont{Hu}},
  \bibinfo{author}{\bibfnamefont{C.}~\bibnamefont{Cheng}},
  \bibinfo{author}{\bibfnamefont{Z.~H.} \bibnamefont{Xu}},
  \bibinfo{author}{\bibfnamefont{H.~G.} \bibnamefont{Luo}}, \bibnamefont{and}
  \bibinfo{author}{\bibfnamefont{S.}~\bibnamefont{Chen}},
  \bibinfo{journal}{Phys. Rev. B} \textbf{\bibinfo{volume}{90}},
  \bibinfo{pages}{035150} (\bibinfo{year}{2014}).

\bibitem[{\citenamefont{Hase et~al.}(2006)\citenamefont{Hase, Kohno, Kitazawa,
  Tsujii, Suzuki, Ozawa, Kido, Imai, and Hu}}]{Hase_PRB_73_104419}
\bibinfo{author}{\bibfnamefont{M.}~\bibnamefont{Hase}},
  \bibinfo{author}{\bibfnamefont{M.}~\bibnamefont{Kohno}},
  \bibinfo{author}{\bibfnamefont{H.}~\bibnamefont{Kitazawa}},
  \bibinfo{author}{\bibfnamefont{N.}~\bibnamefont{Tsujii}},
  \bibinfo{author}{\bibfnamefont{O.}~\bibnamefont{Suzuki}},
  \bibinfo{author}{\bibfnamefont{K.}~\bibnamefont{Ozawa}},
  \bibinfo{author}{\bibfnamefont{G.}~\bibnamefont{Kido}},
  \bibinfo{author}{\bibfnamefont{M.}~\bibnamefont{Imai}}, \bibnamefont{and}
  \bibinfo{author}{\bibfnamefont{X.}~\bibnamefont{Hu}}, \bibinfo{journal}{Phys.
  Rev. B} \textbf{\bibinfo{volume}{73}}, \bibinfo{pages}{104419}
  (\bibinfo{year}{2006}).

\bibitem[{\citenamefont{Hase et~al.}(2007)\citenamefont{Hase, Matsuda, Kakurai,
  Ozawa, Kitazawa, Tsujii, Donni, Kohno, and Hu}}]{Hase_PRB_76_064431}
\bibinfo{author}{\bibfnamefont{M.}~\bibnamefont{Hase}},
  \bibinfo{author}{\bibfnamefont{M.}~\bibnamefont{Matsuda}},
  \bibinfo{author}{\bibfnamefont{K.}~\bibnamefont{Kakurai}},
  \bibinfo{author}{\bibfnamefont{K.}~\bibnamefont{Ozawa}},
  \bibinfo{author}{\bibfnamefont{H.}~\bibnamefont{Kitazawa}},
  \bibinfo{author}{\bibfnamefont{N.}~\bibnamefont{Tsujii}},
  \bibinfo{author}{\bibfnamefont{A.}~\bibnamefont{Donni}},
  \bibinfo{author}{\bibfnamefont{M.}~\bibnamefont{Kohno}}, \bibnamefont{and}
  \bibinfo{author}{\bibfnamefont{X.}~\bibnamefont{Hu}}, \bibinfo{journal}{Phys.
  Rev. B} \textbf{\bibinfo{volume}{76}}, \bibinfo{pages}{064431}
  (\bibinfo{year}{2007}).

\bibitem[{\citenamefont{Bera et~al.}(2022)\citenamefont{Bera, Yusuf, Saha,
  Kumar, Voneshen, Skourski, and Zvyagin}}]{Bera_NC_13_6888}
\bibinfo{author}{\bibfnamefont{A.~K.} \bibnamefont{Bera}},
  \bibinfo{author}{\bibfnamefont{S.~M.} \bibnamefont{Yusuf}},
  \bibinfo{author}{\bibfnamefont{S.~K.} \bibnamefont{Saha}},
  \bibinfo{author}{\bibfnamefont{M.}~\bibnamefont{Kumar}},
  \bibinfo{author}{\bibfnamefont{D.}~\bibnamefont{Voneshen}},
  \bibinfo{author}{\bibfnamefont{Y.}~\bibnamefont{Skourski}}, \bibnamefont{and}
  \bibinfo{author}{\bibfnamefont{S.~A.} \bibnamefont{Zvyagin}},
  \bibinfo{journal}{Nat. Comm.} \textbf{\bibinfo{volume}{13}},
  \bibinfo{pages}{6888} (\bibinfo{year}{2022}).

\bibitem[{\citenamefont{Zeng et~al.}(2022)\citenamefont{Zeng, Song, Ling, Tong,
  Li, Tian, Ma, and Pi}}]{Zeng_CPL_39_107501}
\bibinfo{author}{\bibfnamefont{K.~Y.} \bibnamefont{Zeng}},
  \bibinfo{author}{\bibfnamefont{F.~Y.} \bibnamefont{Song}},
  \bibinfo{author}{\bibfnamefont{L.~S.} \bibnamefont{Ling}},
  \bibinfo{author}{\bibfnamefont{W.}~\bibnamefont{Tong}},
  \bibinfo{author}{\bibfnamefont{S.~L.} \bibnamefont{Li}},
  \bibinfo{author}{\bibfnamefont{Z.~M.} \bibnamefont{Tian}},
  \bibinfo{author}{\bibfnamefont{L.}~\bibnamefont{Ma}}, \bibnamefont{and}
  \bibinfo{author}{\bibfnamefont{L.}~\bibnamefont{Pi}}, \bibinfo{journal}{Chin.
  Phys. Lett.} \textbf{\bibinfo{volume}{39}}, \bibinfo{pages}{107501}
  (\bibinfo{year}{2022}).

\bibitem[{\citenamefont{Higa et~al.}(2017)\citenamefont{Higa, Ding, Yogi,
  Sangeetha, Hedo, Nakama, Onuki, Johnston, and Furukawa}}]{Higa_PRB_96_024405}
\bibinfo{author}{\bibfnamefont{N.}~\bibnamefont{Higa}},
  \bibinfo{author}{\bibfnamefont{Q.~P.} \bibnamefont{Ding}},
  \bibinfo{author}{\bibfnamefont{M.}~\bibnamefont{Yogi}},
  \bibinfo{author}{\bibfnamefont{N.~S.} \bibnamefont{Sangeetha}},
  \bibinfo{author}{\bibfnamefont{M.}~\bibnamefont{Hedo}},
  \bibinfo{author}{\bibfnamefont{T.}~\bibnamefont{Nakama}},
  \bibinfo{author}{\bibfnamefont{Y.}~\bibnamefont{Onuki}},
  \bibinfo{author}{\bibfnamefont{D.~C.} \bibnamefont{Johnston}},
  \bibnamefont{and} \bibinfo{author}{\bibfnamefont{Y.}~\bibnamefont{Furukawa}},
  \bibinfo{journal}{Phys. Rev. B} \textbf{\bibinfo{volume}{96}},
  \bibinfo{pages}{024405} (\bibinfo{year}{2017}).

\bibitem[{\citenamefont{Yasui et~al.}(2014)\citenamefont{Yasui, Kawamura,
  Kobayashi, and Sato}}]{Yasui_JAP_115_17E125}
\bibinfo{author}{\bibfnamefont{Y.}~\bibnamefont{Yasui}},
  \bibinfo{author}{\bibfnamefont{Y.}~\bibnamefont{Kawamura}},
  \bibinfo{author}{\bibfnamefont{Y.}~\bibnamefont{Kobayashi}},
  \bibnamefont{and} \bibinfo{author}{\bibfnamefont{M.}~\bibnamefont{Sato}},
  \bibinfo{journal}{J. Appl. Phys.} \textbf{\bibinfo{volume}{115}},
  \bibinfo{pages}{17E125} (\bibinfo{year}{2014}).

\bibitem[{\citenamefont{Bonner and Fisher}(1964)}]{Bonner_PR_135_A640}
\bibinfo{author}{\bibfnamefont{J.~C.} \bibnamefont{Bonner}} \bibnamefont{and}
  \bibinfo{author}{\bibfnamefont{M.~E.} \bibnamefont{Fisher}},
  \bibinfo{journal}{Phys. Rev.} \textbf{\bibinfo{volume}{135}},
  \bibinfo{pages}{A640} (\bibinfo{year}{1964}).

\bibitem[{\citenamefont{Sachdev}(1994)}]{Sachedev_PRB_50_13006}
\bibinfo{author}{\bibfnamefont{S.}~\bibnamefont{Sachdev}},
  \bibinfo{journal}{Phys. Rev. B} \textbf{\bibinfo{volume}{50}},
  \bibinfo{pages}{13006} (\bibinfo{year}{1994}).

\bibitem[{\citenamefont{Thurber et~al.}(2001)\citenamefont{Thurber, Hunt, Imai,
  and Chou}}]{Thurber_PRL_87_247202}
\bibinfo{author}{\bibfnamefont{K.~R.} \bibnamefont{Thurber}},
  \bibinfo{author}{\bibfnamefont{A.~W.} \bibnamefont{Hunt}},
  \bibinfo{author}{\bibfnamefont{T.}~\bibnamefont{Imai}}, \bibnamefont{and}
  \bibinfo{author}{\bibfnamefont{F.~C.} \bibnamefont{Chou}},
  \bibinfo{journal}{Phys. Rev. Lett.} \textbf{\bibinfo{volume}{87}},
  \bibinfo{pages}{247202} (\bibinfo{year}{2001}).

\bibitem[{\citenamefont{Moriya}(1963)}]{Moriya_JPSJ_18_516}
\bibinfo{author}{\bibfnamefont{T.}~\bibnamefont{Moriya}}, \bibinfo{journal}{J.
  Phys. Soc. Jpn.} \textbf{\bibinfo{volume}{18}}, \bibinfo{pages}{516}
  (\bibinfo{year}{1963}).

\bibitem[{\citenamefont{Gaveau et~al.}(1990)\citenamefont{Gaveau, Boucher,
  Regnault, and Renard}}]{Gaveau_EPL_12_647}
\bibinfo{author}{\bibfnamefont{P.}~\bibnamefont{Gaveau}},
  \bibinfo{author}{\bibfnamefont{J.~P.} \bibnamefont{Boucher}},
  \bibinfo{author}{\bibfnamefont{L.~P.} \bibnamefont{Regnault}},
  \bibnamefont{and} \bibinfo{author}{\bibfnamefont{J.~P.}
  \bibnamefont{Renard}}, \bibinfo{journal}{Europhys. Lett.}
  \textbf{\bibinfo{volume}{12}}, \bibinfo{pages}{647} (\bibinfo{year}{1990}).

\bibitem[{\citenamefont{Shimizu et~al.}(1995)\citenamefont{Shimizu,
  MacLaughlin, Hammel, Thompson, and Cheong}}]{Shimizu_PRB_52_R9835}
\bibinfo{author}{\bibfnamefont{T.}~\bibnamefont{Shimizu}},
  \bibinfo{author}{\bibfnamefont{D.~E.} \bibnamefont{MacLaughlin}},
  \bibinfo{author}{\bibfnamefont{P.~C.} \bibnamefont{Hammel}},
  \bibinfo{author}{\bibfnamefont{J.~D.} \bibnamefont{Thompson}},
  \bibnamefont{and} \bibinfo{author}{\bibfnamefont{S.-W.}
  \bibnamefont{Cheong}}, \bibinfo{journal}{Phys. Rev. B}
  \textbf{\bibinfo{volume}{52}}, \bibinfo{pages}{R9835} (\bibinfo{year}{1995}).

\bibitem[{\citenamefont{Takigawa et~al.}(1995)\citenamefont{Takigawa, Asano,
  Ajiro, and Mekata}}]{Takigawa_PRB_52_R13087}
\bibinfo{author}{\bibfnamefont{M.}~\bibnamefont{Takigawa}},
  \bibinfo{author}{\bibfnamefont{T.}~\bibnamefont{Asano}},
  \bibinfo{author}{\bibfnamefont{Y.}~\bibnamefont{Ajiro}}, \bibnamefont{and}
  \bibinfo{author}{\bibfnamefont{M.}~\bibnamefont{Mekata}},
  \bibinfo{journal}{Phys. Rev. B} \textbf{\bibinfo{volume}{52}},
  \bibinfo{pages}{R13087} (\bibinfo{year}{1995}).

\bibitem[{\citenamefont{Takigawa et~al.}(1996)\citenamefont{Takigawa, Asano,
  Ajiro, Mekata, and Uemura}}]{Takigawa_PRL_76_2173}
\bibinfo{author}{\bibfnamefont{M.}~\bibnamefont{Takigawa}},
  \bibinfo{author}{\bibfnamefont{T.}~\bibnamefont{Asano}},
  \bibinfo{author}{\bibfnamefont{Y.}~\bibnamefont{Ajiro}},
  \bibinfo{author}{\bibfnamefont{M.}~\bibnamefont{Mekata}}, \bibnamefont{and}
  \bibinfo{author}{\bibfnamefont{Y.~J.} \bibnamefont{Uemura}},
  \bibinfo{journal}{Phys. Rev. Lett.} \textbf{\bibinfo{volume}{76}},
  \bibinfo{pages}{2173} (\bibinfo{year}{1996}).

\bibitem[{\citenamefont{Pahari et~al.}(2006)\citenamefont{Pahari, Ghoshray,
  Sarkar, Bandyopadhyay, and Ghoshray}}]{Pahari_PRB_73_012407}
\bibinfo{author}{\bibfnamefont{B.}~\bibnamefont{Pahari}},
  \bibinfo{author}{\bibfnamefont{K.}~\bibnamefont{Ghoshray}},
  \bibinfo{author}{\bibfnamefont{R.}~\bibnamefont{Sarkar}},
  \bibinfo{author}{\bibfnamefont{B.}~\bibnamefont{Bandyopadhyay}},
  \bibnamefont{and} \bibinfo{author}{\bibfnamefont{A.}~\bibnamefont{Ghoshray}},
  \bibinfo{journal}{Phys. Rev. B} \textbf{\bibinfo{volume}{73}},
  \bibinfo{pages}{012407} (\bibinfo{year}{2006}).

\bibitem[{\citenamefont{Jolicoeur and Golinelli}(1994)}]{Jolicoeur_PRB_50_9265}
\bibinfo{author}{\bibfnamefont{T.}~\bibnamefont{Jolicoeur}} \bibnamefont{and}
  \bibinfo{author}{\bibfnamefont{O.}~\bibnamefont{Golinelli}},
  \bibinfo{journal}{Phys. Rev. B} \textbf{\bibinfo{volume}{50}},
  \bibinfo{pages}{9265} (\bibinfo{year}{1994}).

\bibitem[{\citenamefont{Sagi and Affleck}(1996)}]{Sagi_PRB_53_9188}
\bibinfo{author}{\bibfnamefont{J.}~\bibnamefont{Sagi}} \bibnamefont{and}
  \bibinfo{author}{\bibfnamefont{I.}~\bibnamefont{Affleck}},
  \bibinfo{journal}{Phys. Rev. B} \textbf{\bibinfo{volume}{53}},
  \bibinfo{pages}{9188} (\bibinfo{year}{1996}).

\bibitem[{\citenamefont{Sachdev and Damle}(1997)}]{Sachdev_PRL_78_943}
\bibinfo{author}{\bibfnamefont{S.}~\bibnamefont{Sachdev}} \bibnamefont{and}
  \bibinfo{author}{\bibfnamefont{K.}~\bibnamefont{Damle}},
  \bibinfo{journal}{Phys. Rev. Lett.} \textbf{\bibinfo{volume}{78}},
  \bibinfo{pages}{943} (\bibinfo{year}{1997}).

\bibitem[{\citenamefont{Konik}(2003)}]{Konik_PRB_68_104435}
\bibinfo{author}{\bibfnamefont{R.~M.} \bibnamefont{Konik}},
  \bibinfo{journal}{Phys. Rev. B} \textbf{\bibinfo{volume}{68}},
  \bibinfo{pages}{104435} (\bibinfo{year}{2003}).

\bibitem[{\citenamefont{Capponi et~al.}(2019)\citenamefont{Capponi, Dupont,
  Sandvik, and Sengupta}}]{Capponi_PRB_100_094411}
\bibinfo{author}{\bibfnamefont{S.}~\bibnamefont{Capponi}},
  \bibinfo{author}{\bibfnamefont{M.}~\bibnamefont{Dupont}},
  \bibinfo{author}{\bibfnamefont{A.~W.} \bibnamefont{Sandvik}},
  \bibnamefont{and} \bibinfo{author}{\bibfnamefont{P.}~\bibnamefont{Sengupta}},
  \bibinfo{journal}{Phys. Rev. B} \textbf{\bibinfo{volume}{100}},
  \bibinfo{pages}{094411} (\bibinfo{year}{2019}).

\bibitem[{\citenamefont{Fujiwara et~al.}(1992)\citenamefont{Fujiwara, Goto,
  Maegawa, and Kohmoto}}]{Fujiwara_PRB_45_7837}
\bibinfo{author}{\bibfnamefont{N.}~\bibnamefont{Fujiwara}},
  \bibinfo{author}{\bibfnamefont{T.}~\bibnamefont{Goto}},
  \bibinfo{author}{\bibfnamefont{S.}~\bibnamefont{Maegawa}}, \bibnamefont{and}
  \bibinfo{author}{\bibfnamefont{T.}~\bibnamefont{Kohmoto}},
  \bibinfo{journal}{Phys. Rev. B} \textbf{\bibinfo{volume}{45}},
  \bibinfo{pages}{7837} (\bibinfo{year}{1992}).

\bibitem[{\citenamefont{Fujiwara et~al.}(1993)\citenamefont{Fujiwara, Goto,
  Maegawa, and Kohmoto}}]{Fujiwara_PRB_47_11860}
\bibinfo{author}{\bibfnamefont{N.}~\bibnamefont{Fujiwara}},
  \bibinfo{author}{\bibfnamefont{T.}~\bibnamefont{Goto}},
  \bibinfo{author}{\bibfnamefont{S.}~\bibnamefont{Maegawa}}, \bibnamefont{and}
  \bibinfo{author}{\bibfnamefont{T.}~\bibnamefont{Kohmoto}},
  \bibinfo{journal}{Phys. Rev. B} \textbf{\bibinfo{volume}{47}},
  \bibinfo{pages}{11860} (\bibinfo{year}{1993}).

\bibitem[{\citenamefont{Nightingale and Blote}(1986)}]{Nightingale_PRB_33_659}
\bibinfo{author}{\bibfnamefont{M.~P.} \bibnamefont{Nightingale}}
  \bibnamefont{and} \bibinfo{author}{\bibfnamefont{H.~W.~J.}
  \bibnamefont{Blote}}, \bibinfo{journal}{Phys. Rev. B}
  \textbf{\bibinfo{volume}{33}}, \bibinfo{pages}{659} (\bibinfo{year}{1986}).

\end{thebibliography}

\end{document}